\documentclass[prl,twocolumn,amsmath,amssymb,bibnotes,aps,longbibliography]{revtex4-1}
\usepackage{textcomp}
\usepackage{pifont}
\usepackage{mathptmx}
\usepackage{graphicx}
\usepackage{enumerate}
\usepackage{titlesec}
\usepackage{color}
\usepackage[normalem]{ulem}

\usepackage{hyperref}
\hypersetup{
pdftitle={Single-Photon Distillation via a Photonic Parity Measurement Using Cavity QED},
pdfauthor={Severin Daiss, Stephan Welte, Bastian Hacker, Lin Li, and Gerhard Rempe},
colorlinks=true, linkcolor=[RGB]{45,48,146}, citecolor=[RGB]{45,48,146}, filecolor=[RGB]{45,48,146}, urlcolor=[RGB]{45,48,146}}

\newcommand{\bra}[1]{\langle{#1}|}
\newcommand{\ket}[1]{|{#1}\rangle}
\newcommand{\braket}[1]{\langle{#1}\rangle}

\DeclareTextFontCommand{\emph}{\textit}

\makeatletter 
\@ifpackageloaded{hyperref}{\newcommand{\positionlabel}[1]{\hypertarget{#1}{}}}{\newcommand{\positionlabel}[1]{}}
\@ifpackageloaded{hyperref}{}{\newcommand{\hyperref}[2][]{#2}}
\@ifpackageloaded{hyperref}{\newcommand{\figref}[1]{\hyperlink{#1}{\ref*{#1}}}}{\newcommand{\figref}[1]{\ref{#1}}}
\makeatother

\begin{document}
\title{Single-Photon Distillation via a Photonic Parity Measurement Using Cavity QED}

\author{Severin Daiss}
\email{severin.daiss@mpq.mpg.de}
\author{Stephan~Welte}
\author{Bastian~Hacker}
\author{Lin~Li}
\altaffiliation[Present address: ]{School of Physics, Huazhong University of Science and Technology, Wuhan 430074, China}
\author{Gerhard~Rempe}

\affiliation{Max-Planck-Institut f\"ur Quantenoptik, Hans-Kopfermann-Strasse 1, 85748 Garching, Germany}

\begin{abstract}
Single photons with tailored temporal profiles are a vital resource for future quantum networks. Here we distill them out of custom-shaped laser pulses that reflect from a single atom strongly coupled to an optical resonator. A subsequent measurement on the atom is employed to herald a successful distillation. Out of vacuum-dominated light pulses, we create single photons with fidelity $66(1)\%$, two-and-more-photon suppression $95.5(6)\%$, and a Wigner function with negative value $-0.125(6)$. Our scheme applied to state-of-the-art fiber resonators could boost the single-photon fidelity to up to $96\%$.
\end{abstract}

\maketitle \nocite{apsrev41Control}
\textcolor{black}{Pure quantum states are difficult to synthesize in open quantum systems that suffer from environmental perturbations. 
This problem is particularly severe in all research fields that rely on the availability of high-fidelity single-photon states, ranging from quantum communication \cite{kimble2008,scarani2009,reiserer2015} to all-optical quantum computation \cite{knill2001scheme,obrien2007,hacker2016,tiarks2018} and quantum simulation \cite{aaronson2013, spring2013}. Light sources using the fluorescence of a single emitter \cite{brunel1999,kurtsiefer2000,kuhn2002,keller2004,lodahl2015} or a weakly driven nonlinear system \cite{hong1986,duan2001,chou2004, dudin2012strongly, bao2012, peyronel2012quantum, bimbard2014homodyne} have successfully suppressed the emission of two or more photons. However, elimination of the zero-photon component in such sources is much harder to achieve due to conceptual and technical constraints intrinsic to all quantum emitters demonstrated so far. This limits the purity of the produced single-photon state.}

A solution to this problem is distillation that can enhance the fidelity of a target state out of a suitable quantum resource, typically a larger set of states \cite{bennett1996}. Here we employ this idea and filter a single photon out of an incoming light pulse that serves as a photon resource. Instead of using a quantum object as an emitter, we employ it to enhance the single-photon component and suppress both the zero- and the two-photon contributions of the input state. Specifically, we entangle a weak coherent laser pulse with an atom strongly coupled to an optical cavity and then detect the atomic state in a suitable basis. This constitutes a projective parity measurement on the light state and provides a herald that signals suppression of the unwanted vacuum and two-photon components. Our proof-of-concept experiment achieves a single-photon fidelity as high as $66(1)\%$ out of initially vacuum-dominated states. 
Beyond coherent input states, the scheme could be used to block the vacuum contribution from imperfect single-photon sources. Moreover, our protocol is suitable for input states with widely tunable temporal shapes, and can straightforwardly be repeated for further distillation. 
There is no known limit to the achievable single-photon fidelity. 
Additionally, the photonic parity measurement allows one to work with quantum error correcting codes that use the photon parity as an error syndrome \cite{ofek2016}. Our conceptually elegant and technologically simple mechanism is platform independent and can therefore be a useful tool in modular quantum networks with information processing nodes and quantum light sources of the kind described here.
\begin{figure}[b]
\centering
\includegraphics[width=\columnwidth]{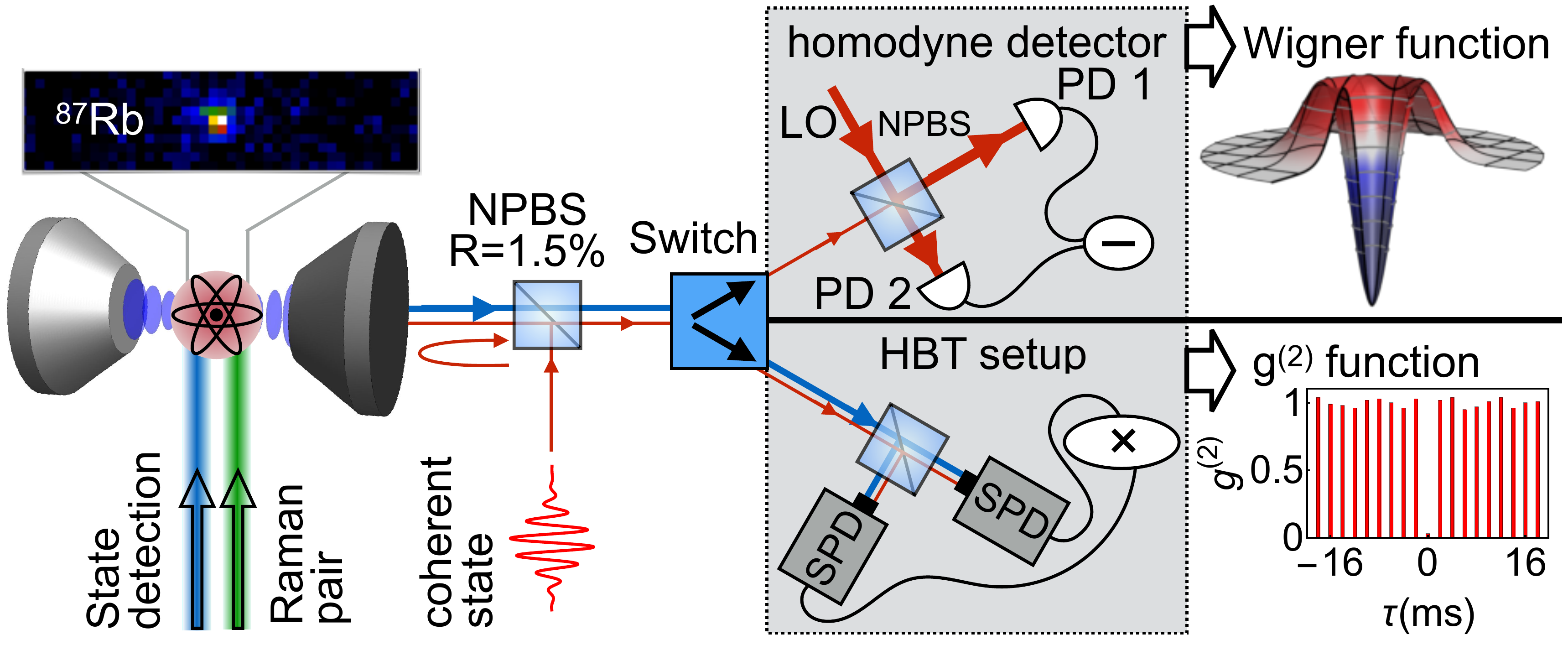}
\caption{\label{fig:setup}Setup of the experiment. Weak coherent pulses are sent to the atom-cavity system using a non-polarizing beam splitter (NPBS). They are reflected from the resonator followed by a state rotation (using a pair of Raman beams) and a state detection of the atom. A switch is used to change between different measurement setups. For homodyne detection, the light is mixed with a local oscillator (LO) beam and detected by two photodiodes (PDs). For coincidence measurements, the light can be sent into a Hanbury Brown--Twiss (HBT) setup, consisting of a beam splitter and two single-photon detectors (SPDs).}
\end{figure}

A schematic of our setup is shown in Fig.~\figref{fig:setup}. We use a single-sided, high-finesse ($6\times10^5$) optical Fabry-P\'{e}rot cavity and trap a single ${}^{87}$Rb atom at its center. The atom can be considered as an effective three-level system consisting of two magnetic hyperfine states of the ground state manifold $\ket{\uparrow}=\ket{5^2S_{1/2},F{=}2,m_F{=}2}$, $\ket{\downarrow}=\ket{5^2S_{1/2},F{=}1,m_F{=}1}$ and an excited state $\ket{e}=\ket{5^2P_{3/2},F{=}3,m_F=3}$. A pair of Raman lasers allows for coherent control of the atomic ground states. The cavity is actively tuned into resonance with the transition $\ket{\uparrow}\leftrightarrow\ket{e}$ at $780\,\mathrm{nm}$. An atom in $\ket{\downarrow}$ is $6.8\,\mathrm{GHz}$ detuned from the resonator such that it effectively does not couple to the cavity mode. The cavity quantum electrodynamics (QED) parameters of our system for an atom occupying the state $\ket{\uparrow}$ are $(g,\kappa,\kappa_r,\gamma)=2\pi(7.8,2.5,2.3,3)\,\mathrm{MHz}$. Here, $g$ is the atom-cavity coupling constant, $\kappa$ and $\kappa_r$ are the cavity-field decay rates in total and through the in-coupling mirror, respectively, and $\gamma$ denotes the atomic polarization decay rate. 

Light coming from the cavity can be directed to two different detection setups. In a homodyne measurement, the light signal is mixed with a resonant local-oscillator beam on a 50:50 beam splitter. Two high-efficiency photodiodes measure the intensity in the two output channels, and their difference in photocurrent is amplified and evaluated by a computer. From measurements with different localoscillator phases, the Wigner function and the density matrix of the light are reconstructed. To prevent backreflected local-oscillator light from entering the cavity, a high-transmission isolator is placed in front of this detection setup. A preceding acousto-optical deflector (AOD) allows us to alternatively send the light from the cavity to a Hanbury Brown--Twiss setup \cite{hanburybrown1956} to analyze its photon statistics by means of another 50:50 beam splitter and two single-photon detectors (SPDs).

The photonic parity measurement presented in this work is based on the reflection of input light from the cavity. We employ resonant coherent pulses described by the state $\ket{\alpha}=e^{-|\alpha|^2/2}\sum_{n=0}^{\infty}(\alpha^n/\sqrt{n!})\ket{n}$, where $\ket{n}$ denotes the respective photonic Fock state and $\alpha$ is assumed to be real. We send pulses with an average photon number $\overline{n}=\alpha^2$, a Gaussian envelope with full width at half maximum $t=2.3\,\mathrm{\mu s}$ and a repetition frequency of $500\,\mathrm{Hz}$ on the in-coupling mirror of the resonator. For a single-photon Fock state $\ket{1}$, a reflection imprints a $\pi$ phase shift on the combined atom-light state for the atom in $\ket{\downarrow}$ as compared to no phase shift for an atom in the coupling state $\ket{\uparrow}$ \cite{duan2004,reiserer2013b}. This can be extended to higher Fock states $\ket{n}$ as long as the intracavity atom is not saturated by the incoming photons. In this regime, a reflection of $\ket{n}$ with the atom in $\ket{\downarrow}$ imprints a phase of $n\pi$ on the combined atom-light state as compared to an atom in $\ket{\uparrow}$. 
Starting with an initial equal superposition $(\ket{\uparrow}+\ket{\downarrow})/\sqrt{2}$, the reflection of Fock states with odd or even photon number turns the atomic state to $(\ket{\uparrow}-\ket{\downarrow})/\sqrt{2}$ or to $(\ket{\uparrow}+\ket{\downarrow})/\sqrt{2}$, respectively. A successive $\pi/2$ rotation maps the state of the atom to $\ket{\uparrow}$ for an odd and to $\ket{\downarrow}$ for an even photon number \cite{wang2005}. The final detection of the atomic state therefore establishes a projective measurement on the photon-number parity of the reflected light. Detecting the atom in $\ket{\uparrow}$ heralds the reflection of odd photon-number components from the initial state and thus a successful distillation: For an input with negligible three- and higher photon components like from an impure photon source or, as applied here, a coherent state $\ket{\alpha}$ with $\alpha^2\lesssim1$, the dominant contribution in the distilled pulse is the single-photon component. Therefore, the protocol can purify a single photon out of an incident classical coherent state in a heralded way.

A versatile tool to analyze the resulting light state is the homodyne detection from which the photonic density matrix $\rho$ can be inferred using a maximum-likelihood approach \cite{lvovsky2009}. All photon-number distributions shown in this work are derived from such a reconstruction. This measurement and the propagation from the cavity to the detector are, of course, subject to losses. From a separate characterization we estimate that they amount to a total of $L_{\textnormal{corr}}=25.1\%$ \cite{supplement}\nocite{kuhn2015,qutip,neuzner_phd}. As they only occur after our distillation scheme, we correct for these losses in the data from the homodyne detection shown in this work.
\begin{figure}
\centering
\includegraphics[width=8.6cm]{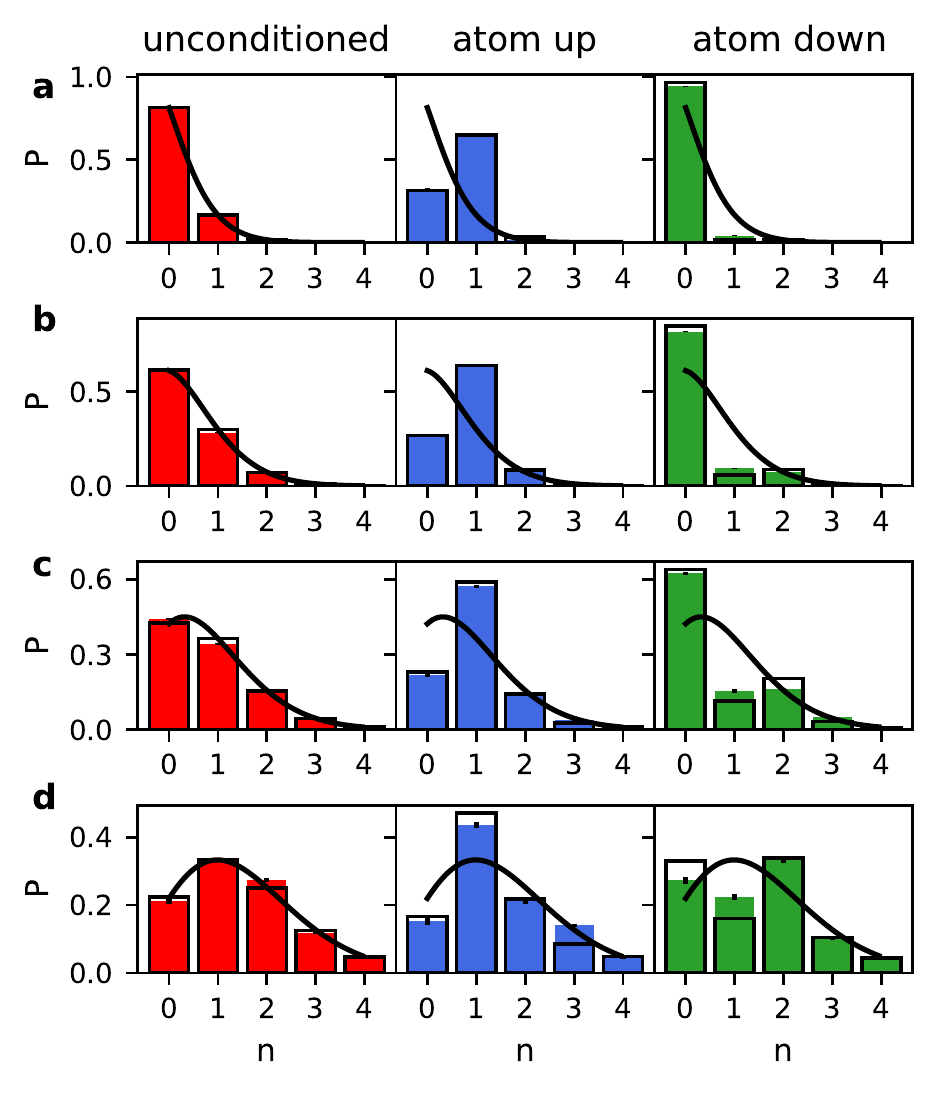}
\caption{\label{fig:photonN}Projecting onto Fock states with different parity. The photon-number distributions of the reflected light unconditioned on a measurement (left), with the atom detected in $\ket{\uparrow}$ (center) and with the atom detected in $\ket{\downarrow}$ (right) are shown. The data are corrected for losses occurring after the cavity. The average photon numbers of the initial coherent pulses are derived from the unconditioned data to be ($0.35$, $0.85$, $1.48$, $2.61$) in (a)-(d). The bars outlined in black indicate the distribution from a calculation with our experimental parameters as discussed in Ref. \cite{supplement}. The Poisson distribution for a coherent state with mean photon number according to the reflected light intensity is shown as a black line.}
\end{figure}
In Fig.~\ref{fig:photonN}, we exemplify the effect of our parity projection by comparing the Fock-state populations derived from the homodyne detection with and without the projecting atomic state detection. The light state unconditioned on the atomic measurement is shown on the left and displays a coherent Poisson distribution. The average photon number is determined by the incoming light pulse and the reflectivity of the cavity \cite{supplement}. Postselecting on a detection of the atom in $\ket\uparrow$ ($\ket{\downarrow}$) projects the light onto odd (even) parity Fock states and the resulting photon-number distributions are shown in the middle (on the right). The coherent distribution can be recovered by adding the conditioned results weighted with the respective heralding probability (discussed below). If the atom is detected in $\ket{\uparrow}$, our protocol strongly reduces the weight of all Fock states with even photon numbers. For small values of $\alpha^2$, this amounts to a state that is dominated by its one-photon contribution. The remaining overlap with the vacuum and with photon states having an even parity results from three different imperfections in the creation of the final light state. The major limitation originates from the parameters of our atom-cavity system. Additional loss modes due to light transmission and scattering leak information about the atomic state to the environment. This leads to a reduction in the contrast of our parity measurement. For an incident weak coherent pulse and an asymmetric cavity, the maximal achievable fidelity of the reflected light with a single-photon state  $F_{\textnormal{1,max}}$ can be calculated from input-output theory \cite{supplement} as 
\begin{equation}
\label{eqn:single_photon_fidelity}
F_{\textnormal{1,max}}=\frac{\kappa_r}{\kappa}\frac{2C}{2C+1},
\end{equation}
with $C=g^2/(2\kappa\gamma)$ being the cooperativity of the atom-resonator system. For our cavity QED parameters, we have $C=4.1$ and a theoretical limit of $F_{\textnormal{1,max}}=0.819$, equivalent to losses of $L_\textnormal{cav}=18.1\%$. Two technical deficiencies reduce the single-photon fidelity of the created light below this fundamental limit. First, small fluctuations of the cavity frequency change the phase difference of the reflected light with the atom in $\ket{\uparrow}$ as compared to $\ket{\downarrow}$.
Second, our atomic state detection yields a wrong result in $1.3\%$ of the cases \cite{supplement}. This leads to an admixture of photonic states of different parity depending on the relative likelihood to herald the atom in $\ket\uparrow$.
\begin{figure}[b!]
\begin{center}
\includegraphics[width=\columnwidth]{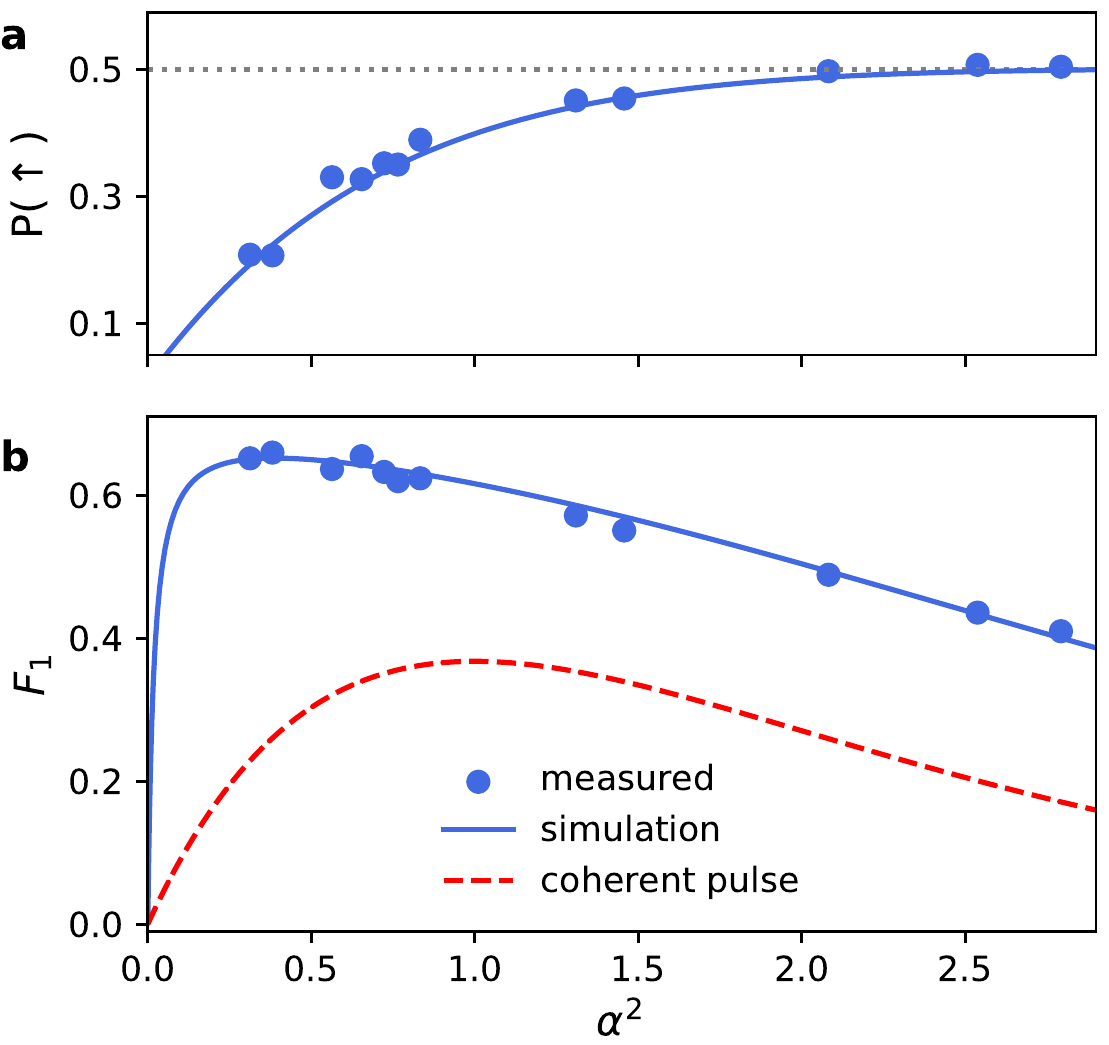}
\end{center}
\caption{\label{fig:fid_eff_single_photon}Projection probability and fidelity with a single photon for different incoming intensities $\alpha^2$. (a)~Heralding probability $P(\uparrow)$ to detect the atom in $\ket{\uparrow}$. (b)~The measured single-photon fidelity $F_1$ of the distilled light is shown as blue data points. The solid line is given by a calculation of our experiment with parameters as described in Ref. \cite{supplement}. For comparison, the overlap of an undistilled coherent state with a single photon is displayed as a dashed red line.}
\end{figure}
The probability for such a detection is shown in Fig.~\ref{fig:fid_eff_single_photon}(a) both measured and calculated for our experimental parameters. For small incoming light intensities, a projection of the light onto odd parity Fock states becomes less likely due to a higher vacuum (even parity) contribution in the initial state. Consequently, the effect caused by the imperfect atomic state detection grows for $\alpha\rightarrow 0$ and results in an increase of the vacuum component. These discussed technical imperfections and possible smaller calibration deficiencies reduce the single-photon fidelites that our protocol achieves for a weak coherent input to $F_1\approx66(1)\%$, with the remaining infidelity being mainly due to the vacuum contribution.

This value is in stark contrast to coherent pulses $\ket{\alpha}$ that are often used as a means to approximate single photons. Such an approach can only yield a fidelity of maximally $F_1=36.8\%$ for a pulse with $\alpha^2=1$, due to additional non-negligible higher Fock-state components. 
In Fig.~\ref{fig:fid_eff_single_photon}(b), we compare the single-photon fidelity $F_1$ of a weak coherent pulse to the data our scheme achieves. The light after the distillation clearly shows strongly improved single-photon contributions above the maximum $F_1$ for coherent light over the full range of considered incoming intensities $\alpha^2$.
\begin{figure}[t]
\centering
\includegraphics[width=8.6cm]{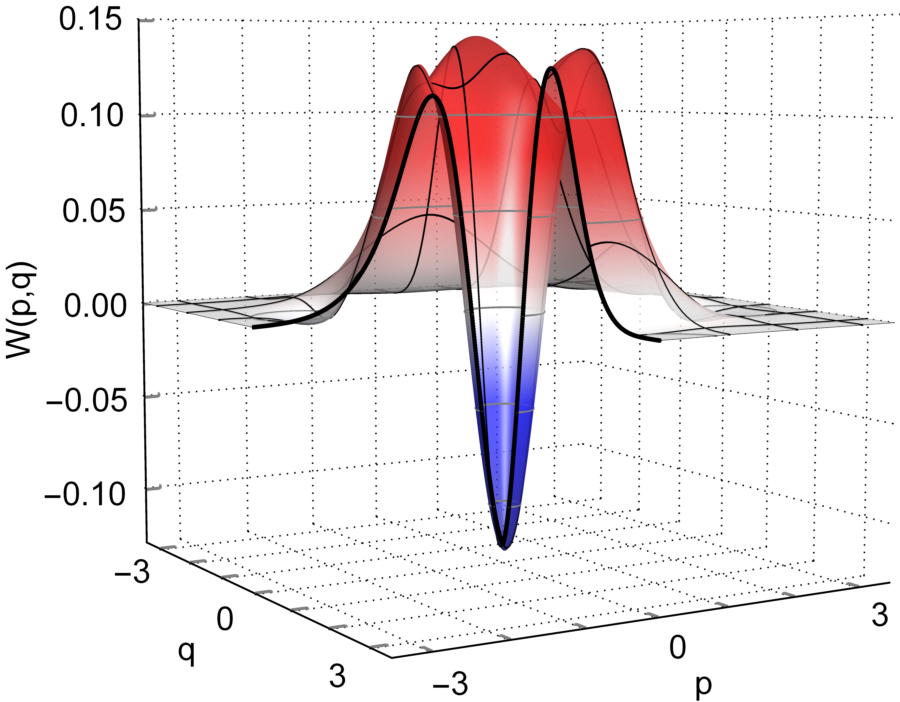}
\caption{
\label{fig:wigner_corrected}Single-photon Wigner function. The plot shows the reconstructed Wigner function of the produced single-photon state for an incoming coherent state with $\alpha^2=0.31$. The two field quadratures spanning phase space are labeled $p$ and $q$. The data are corrected for propagation and detection losses of $25.1\%$.}
\end{figure}

From the homodyne measurement, we can furthermore reconstruct the Wigner function of the final light state. Additionally to the the diagonal elements of the density matrix discussed so far, it also contains information about the coherences. We show the reconstruction for incoming coherent light of $\alpha^2=0.31$ in Fig.~\ref{fig:wigner_corrected}. The Wigner function exhibits the general form expected for a single photon and we observe negativity with a minimum value of $-0.125(6)$. This verifies the significant nonclassicality of the produced state. The data shown are corrected for losses in propagation and detection. The uncorrected reconstruction including the effects that occur after a successful state distillation is given in the Supplemental Material \cite{supplement}.

\begin{figure}[t!]
\centering
\includegraphics[width=8.6cm]{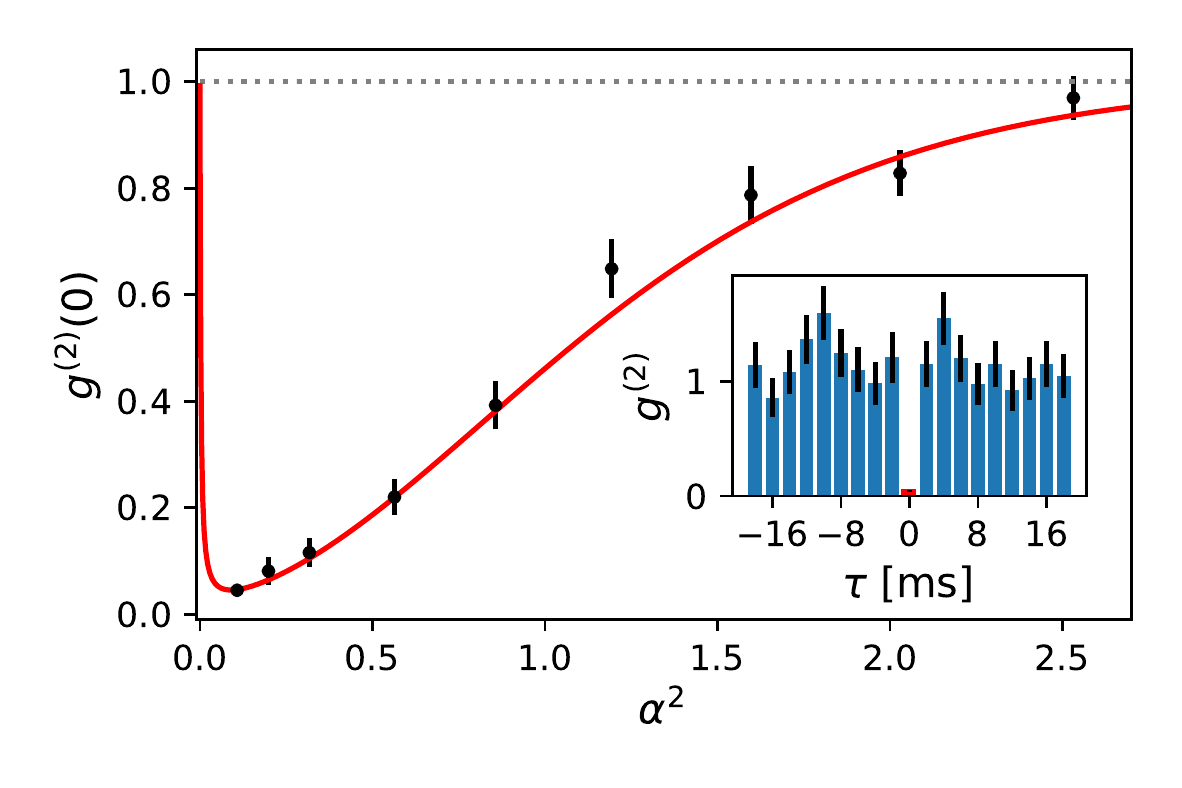}
\caption{\label{fig:g2}Second-order correlation function.
The measured data points show $g^{(2)}(0)$ of the light after a heralding state-detection event as a function of the incoming intensity $\alpha^2$. The solid line is a calculation of the expected $g^{(2)}(0)$ for our system including dark counts of the SPDs (see Ref. \cite{supplement} for details). In the inset of the figure, $g^{(2)}(\tau)$ for $\alpha^2=0.11$ is displayed.}
\end{figure}

For incoming light intensities that are smaller, the imperfect state detection leads to an increasing admixture of even parity photon numbers and therefore to a vacuum-dominated measurement in the homodyne setup. To quantify the suppression of higher photon contributions that our protocol achieves for an incident $\ket{\alpha}$ in a loss independent way, we analyze the distilled light with the Hanbury Brown--Twiss setup. By normalizing to the click probability of the used detectors individually, the effect of the vacuum is excluded in such an analysis.
We measure the second-order correlation function $g^{(2)}(\tau)$ of the purified pulse for different incoming intensities between $\alpha^2=0.1$ and $\alpha^2=2.5$. Results are shown in Fig.~\ref{fig:g2}. They display sub-Poissonian photon statistics indicated by reduced intensity fluctuations $g^{(2)}(0)<1$. For $\alpha^2\rightarrow 0$, the measurements suggest $g^{(2)}(0)\rightarrow 0$ as anticipated for vanishing higher Fock-state contributions.
Figure \ref{fig:g2} also shows a calculated curve of the expected $g^{(2)}(0)$ for our system including the dark-count rate of the used SPDs. In the limit of few detections for small $\alpha^2$, these dark counts of the SPDs result in an uncorrelated signal that ultimately limits the lowest value obtained for $g^{(2)}(0)$. 
For $\alpha^2=0.11$, we evaluate the correlations between light pulses in different experimental runs (inset in Fig.~\ref{fig:g2}). The small value of $g^{(2)}(0)=0.045(6)$ is accompanied by $g^{(2)}(\tau\neq 0)\approx1$. Therefore we observe over the time interval of one distilled pulse an antibunching of photons. This together with $g^{(2)}(0)\rightarrow0$ for weak incident light confirms the suppression of the two-photon component of an initial state.

The temporal shape of the distilled light is determined by the incoming pulse. This is in contrast to e.g.\ the application of adiabatic transfer methods to generate single photons: Producing specific photon shapes with such a scheme requires the inversion of nontrivial functions and a precise control of the intensity and phase of a coupling laser \cite{gorshkov2007b}. The target temporal form in this case is therefore necessarily predefined. Our protocol, however, allows us to quickly change between different shapes, e.g., to counteract distortions in different channels of a quantum network. Furthermore, it is not \textit{a priori} necessary to know the final form. Although the experiments discussed before were exclusively performed with coherent pulses having a Gaussian envelope, our protocol is nevertheless robust as long as the spectral width of the photon is smaller than the cavity linewidth $\kappa$. We exemplify this by measuring $g^{(2)}(0)$ for two additional temporal pulse shapes specified in Ref. \cite{supplement} with $\alpha^2=0.11$. Table \ref{tab:g2shapes} lists the results obtained for the correlation function. We observe values of $g^{(2)}(0)=0.06(3)$ for both newly considered shapes, which agrees with the results obtained for the Gaussian pulse.
\begin{table}[b]
\caption{\label{tab:g2shapes}%
Second-order correlation function for different photon shapes. We list the measured values for $g^{(2)}(0)$ for temporal profiles with $\alpha^2=0.11$. 
}
\begin{ruledtabular}%
\begin{tabular}{lc}%
Photon shape & $g^{(2)}(0)$ \\
\hline
Gaussian & $0.045(6)$\\
Double peak & $0.06(3)$\\
Rectangular & $0.06(3)$\\
\end{tabular}%
\end{ruledtabular}%
\end{table}

While we demonstrated the performance of our protocol with incident weak coherent states, it is worthwhile to analyze the effect our distillation would have on an impure single-photon state. As a reference, we choose a light state produced in a similar system using an adiabatic transfer method \cite{muecke2013}, where single-photon fidelities of $F_1=56\%$ have been measured. Applying our protocol with all limiting technical imperfections to the corresponding state, we could directly boost this fidelity to $F_1=70.1\%$ with an overall probability of $45.2\%$ for a heralding event. Our system can be further improved with a better cavity stabilization and an enhanced atomic-state detection. The maximum fidelity is then only limited by loss channels of the cavity QED system and can be calculated according to Eq.~(\ref{eqn:single_photon_fidelity}) to $F_{\textnormal{1,max}}=81.9\%$ for our parameters. There is no limit to improve this value, e.g., with better mirrors and a smaller mode volume. Indeed, in state-of-the-art fiber resonators, atom-cavity coupling constants of $g=2\pi\times240\,\mathrm{MHz}$ have been achieved for a single ${}^{87}$Rb atom in a cavity of length $39\,\mathrm{\mu m}$ \cite{gehr2010}. Combining this value with realistic parasitic mirror losses of $13.5\,\mathrm{ppm}$ \cite{uphoff2015} and an out-coupling transmission of $1300\,\mathrm{ppm}$, single-photon fidelities of $F_{\textnormal{1,max}}=95.9\%$ can be realized.

The availability of a herald in our scheme allows for a repeat-until-success or for a parallel mode of operation of several such photon sources as a means to enhance the efficiency. In addition, the sensitivity of our protocol to the photon-number parity can be employed to purify the Fock-state distribution, e.g., of optical cat states \cite{ourjoumtsev2007,hacker2019} or light states generated by photon-blockade mechanisms \cite{hamsen2017,loredo2018}. For example, the even photon-number components of a given distribution could be suppressed or enhanced by postselecting on the corresponding heralding event. All of this makes our distillation scheme a promising tool for future quantum networks.
\begin{acknowledgments}
This work was supported by the Deutsche Forschungsgemeinschaft via the excellence cluster Nanosystems Initiative Munich (NIM) and the European Union via the flagship project Quantum Internet Alliance (QIA). S.W.\ was supported by Elitenetzwerk Bayern (ENB) through the doctoral program Exploring Quantum Matter (ExQM).
\end{acknowledgments}

\section{SUPPLEMENTAL MATERIAL}
\label{supplement}

\section{I. Theoretical Considerations}
\subsection{Maximally Achieveable Overlap with a Single Photon}  
Here we consider the theory of our distillation operation by employing cavity input-output theory \cite{kuhn2015}. 
We take four light modes after the interaction with the cavity into account: The reflection from the resonator $\ket{r}$, the transmission through it $\ket{t}$, and the loss modes for scattering from the atom $\ket{a}$ and from the mirrors $\ket{m}$. For a coherent input $\ket{\alpha}$ with a real valued amplitude $\alpha$, the light in these modes can also be considered as coherent with amplitudes that depend on the number of coupling atoms $N=1$ for $\ket{\uparrow}$ or $N=0$ for $\ket{\downarrow}$ and are given by
\begin{eqnarray}
r_{\downarrow/\uparrow} &=& \frac{Ng^2+(\left(i\Delta_c+\kappa\right)-2\kappa_r)\left(i\Delta_a+\gamma\right)}{Ng^2+\left(i\Delta_c+\kappa\right)\left(i\Delta_a+\gamma\right)}\alpha,
\label{supp_eqn_inputoutput_1}
\\
t_{\downarrow/\uparrow} &=& \frac{2\sqrt{\kappa_r\kappa_t}\left(i\Delta_a + \gamma\right)}{Ng^2+\left(i\Delta_c+\kappa\right)\left(i\Delta_a+\gamma\right)}\alpha,
\\
m_{\downarrow/\uparrow} &=& \frac{2\sqrt{\kappa_r\kappa_m}\left(i\Delta_a + \gamma\right)}{Ng^2+\left(i\Delta_c+\kappa\right)\left(i\Delta_a+\gamma\right)}\alpha,
\\
a_{\downarrow/\uparrow} &=& \frac{2\sqrt{\kappa_r\gamma}\sqrt{N}g}{Ng^2+\left(i\Delta_c+\kappa\right)\left(i\Delta_a+\gamma\right)}\alpha
.
\label{supp_eqn_inputoutput_4}
\end{eqnarray}
Here, $\kappa_r$,  $\kappa_t$ and $\kappa_m$ are the decay rates of the cavity field into the reflection mode, into the transmission mode, and into modes associated with the scattering on the mirrors, respectively. $\Delta_a$ and $\Delta_c$ give the detunings of the incoming light to the atom and to the cavity resonance.
We assume an asymmetric cavity with a non-zero reflectivity and label the light in the reflected mode $\ket{r_{\uparrow}}$ or $\ket{r_{\downarrow}}$ depending on the atomic state. Similarly, we summarize the transmission and scattering losses with a single loss mode  $\ket{l_{\uparrow}}:=\ket{t_{\uparrow}}\ket{m_{\uparrow}}\ket{a_{\uparrow}}$ and $\ket{l_{\downarrow}}:=\ket{t_{\downarrow}}\ket{m_{\downarrow}}\ket{a_{\downarrow}}$. The overlaps of these different modes can be calculated using the amplitudes given by Eq.~\ref{supp_eqn_inputoutput_1}--\ref{supp_eqn_inputoutput_4} and amount to
\begin{eqnarray}
&\braket{r_\uparrow|r_\downarrow}& = e^{-2\xi^2\alpha^2},
\\
&\braket{l_\uparrow|l_\downarrow}& = \braket{t_\uparrow|t_\downarrow}\braket{m_\uparrow|m_\downarrow}\braket{a_\uparrow|a_\downarrow} = e^{-2(1-\xi)\xi\alpha^2},
\\
\bra{l_\uparrow}&\braket{r_\uparrow|r_\downarrow}&\ket{l_\downarrow} = e^{-2\xi\alpha^2},
\end{eqnarray}
where we introduce the shorthand \begin{equation}
    \xi:=\frac{\kappa_r}{\kappa}\frac{g^2}{g^2+\kappa\gamma}=\frac{\kappa_r}{\kappa}\frac{2C}{2C+1},
\end{equation}
with the cooperativity $C=g^2/(2\kappa\gamma)$ of the atom-resonator system. For our cavity parameters, we find $C=4.1$ and $\xi=0.819$. 
Reflecting coherent light from the cavity with an atom in an equal superposition state $(\ket{\uparrow}+\ket{\downarrow})/\sqrt{2}$, performing a subsequent $\pi/2$ rotation and detecting the atomic state, the resulting photonic state is
\begin{equation}
\ket{\psi^\mp_\textnormal{out}} = \frac{\ket{r_\uparrow}\ket{l_\uparrow}\mp\ket{r_\downarrow}\ket{l_\downarrow}}{\sqrt{2(1\mp e^{-2\xi\alpha^2})}}.
\end{equation}
The $-$ ($+$) corresponds to a state-detection result of $\ket{\uparrow}$ ($\ket{\downarrow}$) and therefore to light that is projected on the odd (even) Fock photon-number states. 

As the light in the loss modes is lost to the environment, we trace out these degrees of freedom to arrive at the density matrix corresponding to the detected light 
\begin{equation}\begin{split}
\rho^\mp =& \operatorname{tr}_{l} \ket{\psi_\textnormal{out}^\mp}\bra{\psi_\textnormal{out}^\mp} \\
=& \frac{
  \ket{r_\uparrow}\bra{r_\uparrow}
\mp e^{-2(1-\xi)\xi\alpha^2}(\ket{r_\uparrow}\bra{r_\downarrow}
+ \text{c.t.})
+ \ket{r_\downarrow}\bra{r_\downarrow}
}{2(1\mp e^{-2\xi\alpha^2})}.
\end{split}
\label{supp_eqn_rho}
\end{equation}
The loss of information leads to a damping of the coherence terms in the density matrix. This is equivalent to a pure state that was subjected to losses amounting to $L=1-\xi$. Distilling a single photon out of a coherent input, the parasitic cavity modes establish the achievable limit of the fidelity $F_1$. For $\alpha^2\rightarrow0$ and $L\le1/\sqrt{3}$, it is bounded by the resonator losses to
\begin{equation}
F_{1}\le\xi=\frac{\kappa_r}{\kappa}\frac{2C}{2C+1}.
\end{equation}
For our cavity parameters $F_{1}\le0.819$. This is a measure for the quality of our protocol to distill single photons out of an initial weak coherent pulse, where the only limitation is due to our cavity parameters. We can calculate the overlap of the resulting density matrix for incoming pulses of different intensities with a Fock state $\ket{n}$ as
\begin{equation}\begin{split}
\braket{n|\rho^-|n}&=\frac{1}{2(1-e^{-2\xi\alpha^2})}\left(
e^{-|r_\uparrow|^2}\frac{(r_\uparrow)^{2n}}{n!}\right.\\&-2e^{-2(1-\xi)\xi\alpha^2}e^{-\frac{1}{2}\left(|r_\uparrow|^2+|r_\downarrow|^2\right)}\frac{(r_\uparrow r_\downarrow)^n}{n!} \\&+\left. e^{-|r_\downarrow|^2}\frac{(r_\downarrow)^{2n}}{n!}
\right).
\label{supp_eqn_overlap}
\end{split}
\end{equation}
While this describes the fidelities for the ideal state achieveable with our cavity, there are additional imperfections in our setup that amount to further losses in the relative intensity and that are discussed in the next section.

\subsection{Errors and Losses Compared to an Ideal Single Photon}
We first discuss the general effect of additional losses on the density matrix of the produced light state before we detail various imperfections of our experimental setup.

Acting with losses $L$ on the density matrix of Eq.~(\ref{supp_eqn_rho}), it evolves to
\begin{equation}\begin{split}
\rho^\mp =& \frac{1}{2(1\mp e^{-2\xi\alpha^2})}\left(\ket{\nu r_\uparrow}\bra{\nu r_\uparrow}+\ket{\nu r_\downarrow}\bra{\nu r_\downarrow}\right.\\
&\mp\left. e^{-2L\xi^2\alpha^2}e^{-2(1-\xi)\xi\alpha^2}(\ket{\nu r_\uparrow}\bra{\nu r_\downarrow}
+ \text{c.t.}) \right),
\end{split}
\end{equation}
where we have used $\nu = \sqrt{T}$ with the transmission $T=1-L$. The overlap with a Fock state $\ket{n}$ from Eq.~\ref{supp_eqn_overlap} becomes
\begin{equation}\begin{split}
\braket{n|\rho^-|n}&=\frac{T^n}{2(1-e^{-2\xi\alpha^2})}\left(
e^{-T|r_\uparrow|^2}\frac{(r_\uparrow)^{2n}}{n!}\right.\\&-2e^{-2(1-\xi)\xi\alpha^2}e^{-2L(\xi\alpha)^2}e^{-\frac{1}{2}T\left(|r_\uparrow|^2+|r_\downarrow|^2\right)}\\&\cdot\frac{(r_\uparrow r_\downarrow)^n}{n!}+\left. e^{T|r_\downarrow|^2}\frac{(r_\downarrow)^{2n}}{n!}
\right).
\end{split}
\end{equation}
Especially for small $\alpha^2$, the losses lead to an increased admixture of vacuum in the resulting state.

To arrive at the photonic states that we measure with our homodyne detection, we distinguish two different types of additional imperfections. The first kind affects the propagation of the light and its subsequent detection. The use of various optical elements, an acousto-optical deflector and an isolator leads to transmission losses of the light. In the homodyne detection setup, there are additionally imperfect mode matching to the local oscillator and intensity fluctuations thereof, electronic noise, filtering and the non-unity quantum efficiency of the used photodiodes. We summarize the estimated lower bounds of these losses in Table~\ref{tab:losses} and arrive at total of $L=25.1\%$. As these losses are imposed on the light state after its production, we correct for them in the data shown in the main text.
\renewcommand{\thetable}{SI}
\begin{table}[t]
\caption{\label{tab:losses}
Loss budget listing the loss channels in propagation and detection that are not part of the production process. Unless otherwise specified, the uncertainties on the given values are on the order of $10\%$. Individually listed losses $i$ are summed according to $1-L_\textnormal{sum}=\prod_i(1-L_{i})$. 
}
\begin{ruledtabular}%
\begin{tabular}{lr}%
Source of loss & \hspace{-2em}(effective) Loss $L_i$ \\
\hline
Other optics: waveplates, mirrors and NPBS & 7.6\% \\
Mode matching with local oscillator (both paths) & 6.0\% \\
Limited isolator transmission & 3.0\% \\
Switch acousto-optical deflector & 2.5\% \\
Detector dark noise & 2.5\% \\
Laser classical noise & 1.8\% \\
NPBS reflectivity & 1.5\% \\
Limited quantum efficiency of homodyne detector & $1.5(15)\%$\\
Electronic high pass 0.7\,kHz signal reduction & 1.1\% \\
Vacuum viewport reflection & 0.6\% \\
Electronic high pass background noise & 0.2\% \\
\hline
Total losses $L_\textnormal{sum}$ & 25.1\%
\end{tabular}%
\end{ruledtabular}%
\end{table}

The second kind of imperfections affect the production of the light state itself and therefore we do not correct for them. The main effects are fluctuations of the cavity resonance frequency and an imperfect state detection of the atom. The first results mainly in a variation of the phase-difference between $\ket{r_\uparrow}$ and $\ket{r_\downarrow}$ that reduces the contrast of the parity projection. As for the atomic state detection, we infer that we obtain the wrong result in $1.3\%$ of all detection events (discussed below). For an incident coherent pulse, the resulting effect depends on the average photon number via the probabilities to detect the atom in the state $\ket\uparrow$ or $\ket\downarrow$: We have $\braket{\alpha|\rho^-|\alpha}\rightarrow 0$ and $\braket{\alpha|\rho^+|\alpha}\rightarrow 1$ for $\alpha^2\rightarrow 0$. This leads to a bigger admixture of wrong detection events for small $\alpha$ when postselecting on $\ket\uparrow$. The total resulting density matrix can be calculated using Bayes' theorem as
\begin{equation}
\rho_\textnormal{eff} = \left(0.987\braket{\alpha|\rho^-|\alpha} \rho^- +
0.013\braket{\alpha|\rho^+|\alpha} \rho^+\right)/\textnormal{P}(\uparrow),
\end{equation}
where $\textnormal{P}(\uparrow)$ is the total probability to detect the atom in state $\ket{\uparrow}$. 
Lastly, there are additional losses that are not accounted for in the loss budget as outlined above, either because we underestimate existing loss channels or because imperfect calibration of the experiment introduces further effects. 
\subsection{Parameters to for Calculations on the Experiment}
We calculate the theory of the photon parity measurement in python using Qutip \cite{qutip} and implement the model discussed above for our cavity parameters $g$, $\kappa$, $\kappa_r$ and $\gamma$.
To quantify the amount of losses, we fit the calculation to the measured populations in the Fock states $\ket{0}$, $\ket{1}$ and $\ket{2}$ as derived from the homodyne detection for different incoming intensities $\alpha^2$.
We use three free parameters namely the total losses on the light $L_{\textnormal{fit}}$ compared to the ideal state for our cavity, the fraction of an atomic state detection yielding the wrong result and the detuning of the light to the cavity $\Delta_c$.
The atomic resonance is AC-Stark shifted by residual atomic motion in the standing-wave dipole trap \cite{neuzner_phd}, which amounts to $\Delta_a=6\,\mathrm{MHz}$ on average.
Using normal-mode spectroscopy, we tune the shifted transition frequency into resonance with the incoming light.
Both $\Delta_a$ and $\Delta_c$ are subject to slow uncontrolled drifts by a few hundred kilohertz due to temperature changes of the atom as well as the cavity mirror substrate.
The fit yields $L_{\textnormal{fit}}=35.2\%$, an erroneous state detection of $1.3\%$ and $\Delta_c=0.39\,\mathrm{MHz}$.
In all calculations for the homodyne data in the main text, these values are used and a correction for the transmission and propagation of $L_\textnormal{sum}=25.1\%$ is applied. The remaining losses of $L_{\textnormal{uncorr}} = 1-(1-L_\textnormal{fit}) / (1-L_\textnormal{sum}) = 13.5\%$ persist in all presented data.
For the calculation of the $g^{(2)}$ correlation function, we use the parameters given here for the wrong state detection and $\Delta_a$. In this experiment however, we actively stabilize $\Delta_c$ to zero via active feedback to the cavity lock laser.
Furthermore, we use the separately characterized dark-count rate of $20\,\mathrm{Hz}$ of the two single-photon detectors in the Hanbury Brown--Twiss setup.

\section{II. Experimental Details}
\subsection{Homodyne Detection}
Here we specify some additional details of our homodyne setup. The light coming from the cavity is mixed on a 50:50 non-polarizing beam splitter with coherent light of $1.8\,\mathrm{mW}$ derived from the same laser source. The intensity in the output channels is measured with a home-built homodyne detector. It employs two Hamamatsu photodiodes of type S3883 with removed glass windows that have a quantum efficiency of $\eta=98.5(15)\%$. Their photocurrent difference is amplified, digitized, recorded by a field-programmable gate array (FPGA) and then sent to a computer for data evaluation. To prevent back scattering of light from the strong local oscillator by the elements in the homodyne detector, we use an optical isolator (Toptica, model SSR780) with a transmission of $97\%$. 

\renewcommand{\thefigure}{S1}
\begin{figure}[b!]
\centering
\includegraphics[width=8.6cm]{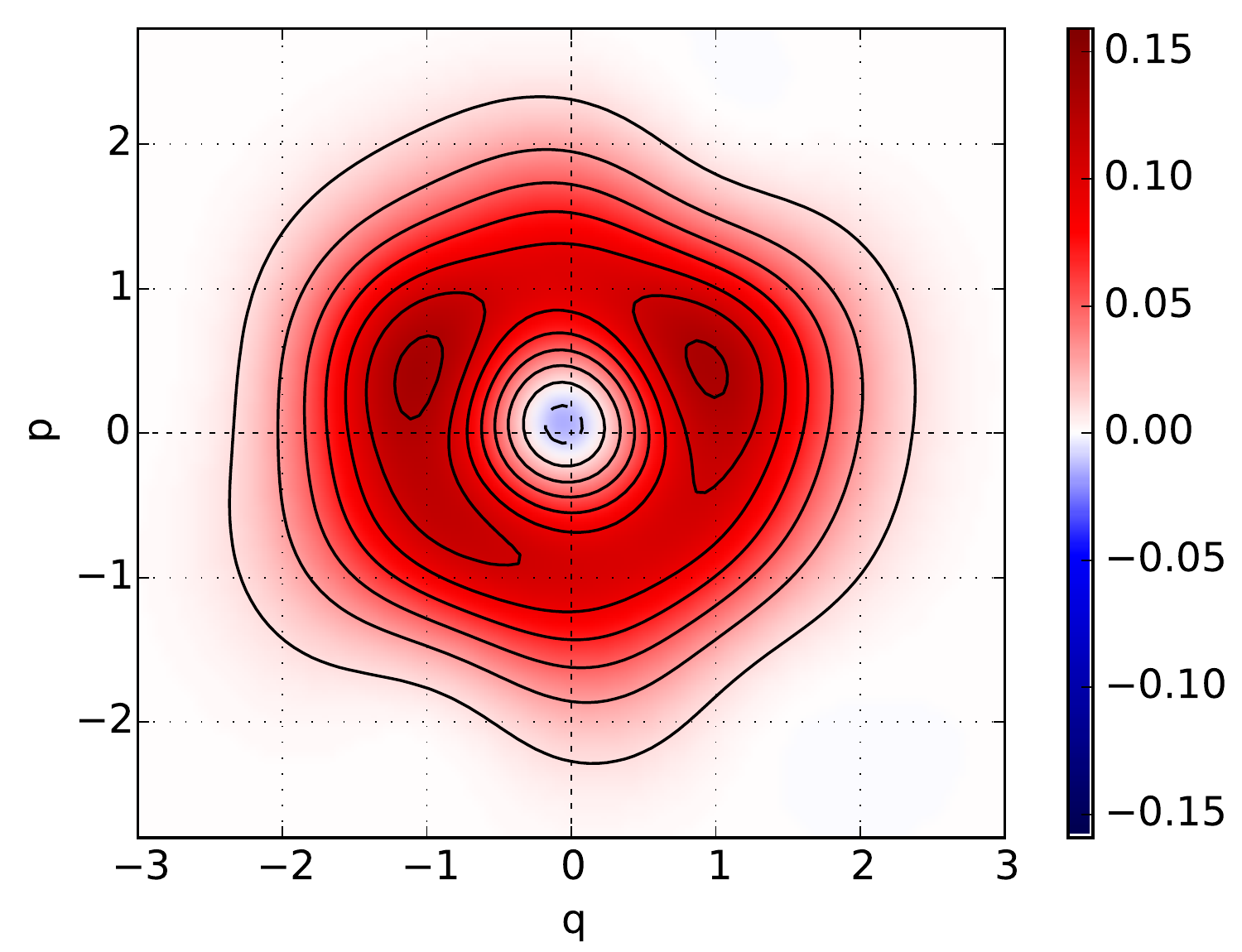}
\caption{\label{fig:wigner_uncorrected}
Uncorrected Wigner function.
We observe a minimum value of $-0.016(8)$ in the reconstructed Wigner function.
A slight asymmetry in the distribution is caused by residual higher Fock-state contributions for the incoming intensity of $\alpha^2=0.31$ and the additional effect of losses.
Correcting this data for the propagation and detection losses after the cavity results in the Wigner function shown in Fig.~4 in the main text.}
\end{figure}
\subsection{Uncorrected Wigner Function}
In the main text, we present a reconstructed Wigner function in Fig.~4. The data are corrected for transmission and detection losses that occur after the distillation process. In Fig.~\ref{fig:wigner_uncorrected}, we show a reconstruction of the same data set, which still includes those losses.
The negativity in this uncorrected Wigner function is less pronounced than in the main text due to the additional losses and we have a minimum value of $-0.016(8)$. 

\subsection{Different Tested Pulse Shapes}
To demonstrate that our protocol works reliably for various pulse shapes, we measure the $g^{(2)}$ function for different temporal profiles. They are created by adjusting the incoming coherent state with an acousto-optical modulator. Fig.~\ref{fig:pulseshape} shows the intensity profiles of the applied photon pulses as derived from the clicks of the single-photon detectors. The corresponding values for $g^{(2)}(0)$ are given in the main text in Table~I. The average photon number for all different pulse forms is $\alpha^2=0.11$.
\renewcommand{\thefigure}{S2}
\begin{figure}[h]
\centering
\includegraphics[width=8.6cm]{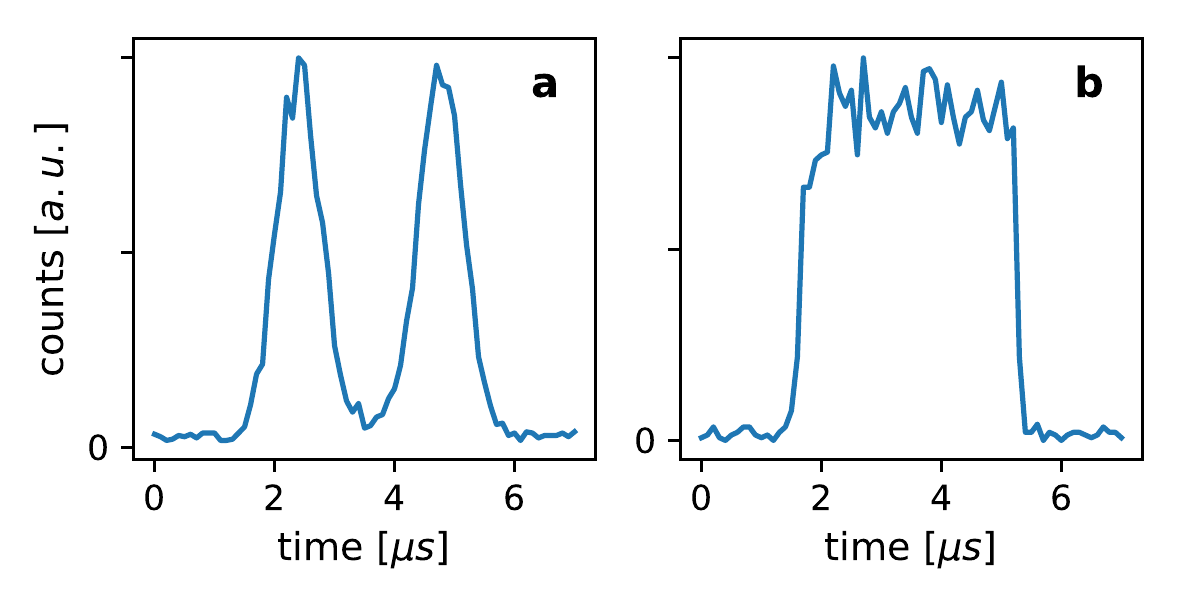}
\caption{\label{fig:pulseshape}Different tested pulse shapes evaluated with single-photon detector clicks. The vertical resolution is shot-noise limited. (a)~Double peak. (b)~Rectangular pulse.}
\end{figure}

\clearpage
\end{document}